\documentclass{iopjournal}
\usepackage[T1]{fontenc} 
\usepackage{booktabs}
\usepackage{subcaption}
\usepackage{xr}
\usepackage{ulem}
\usepackage{lineno}

\usepackage{graphicx,psfrag,amsmath,amssymb,amsfonts,latexsym,color,dcolumn,bm}
\usepackage{soul} 

\begin{document}
\articletype{Paper} %
\title{Near-field-driven Radiative Thermal Dynamics in Aperiodic Photonic Nanostructures}

\author{M. Prado$^1$, A. Manjavacas$^2$, F.A. Pinheiro$^{1,*}$, W.J.M. Kort-Kamp$^{3,*}$}\\

\affil{$^1$ Instituto de F\'{\i}sica, Universidade Federal 
Rio de Janeiro, 21941-972, Rio de Janeiro, RJ, Brazil}\\

\affil{$^2$ Instituto de Qu\'{\i}mica F\'{\i}sica Blas Cabrera (IQF), CSIC, 28006 Madrid, Spain}\\

\affil{$^3$ Theoretical Division, Los Alamos National Laboratory,
MS B262, Los Alamos, New Mexico 87545, USA}\\

\email{fpinheiro@if.ufrj.br, kortkamp@lanl.gov}





\begin{abstract}

Harnessing structural correlations in near-field plasmonic and polaritonic systems hold untapped potential for controlling light–matter interactions at the nanoscale. By tuning these correlations, one can reshape mode localization, coupling, and spectral distribution which are properties central to manipulating energy transport and field enhancement in nanophotonic platforms. We exploit Vogel spirals, an aperiodic geometry where a single parameter dictates spatial correlations, to show how correlation strength reshapes the modal spectrum and transient dynamics of near-field coupling. As a proof of concept, we demonstrate that aperiodic configurations outperform both uncorrelated (random) and periodic arrays in controlling near-field radiative heat-transfer dynamics. These results establish deterministic aperiodic order as a powerful platform for tailoring correlated electromagnetic responses in next-generation nanophotonic devices.

\end{abstract}

\pagebreak 

\section{\label{sec:sec1} Introduction}
Tailoring structural correlations is fundamental to unlocking novel optical functionalities and advancing the design of next-generation photonic technologies. These correlations have a profound impact on light scattering, diffraction, transport, and localization \cite{ishimaru1978wave, carminati2021principles}. At the microscopic level, they govern complex near-field interactions and interference effects that ultimately shape the macroscopic transport behavior of electromagnetic waves. 

Among the most versatile platforms for engineering such correlations are deterministic aperiodic systems, metamaterials characterized by a tunable degree of aperiodic structural order derived from deterministic mathematical rules. These systems have emerged as powerful tools for manipulating light-matter interactions, offering unprecedented control over structural correlations and enabling functionalities beyond those available in traditional optical materials \cite{Gopinath,Lee,noh2011lasing,DalNegroReview,dal2022waves,MaciaBook,MaciaBook2,DalNegroCrystals,sgrignuoli2020multifractality,sgrignuoli2020subdiffusive,TPSE}. In fact, unlike photonic crystals \cite{john1987strong,yablonovitch1987inhibited}, fully disordered \cite{wiersma2013disordered}, or correlated disordered optical media \cite{yu2021engineered,vynck2023light, salvatore2022, salvatore2024, salvatore2025}, deterministic aperiodic systems support distinct and rich optical phenomena, including fractal transmission spectra, light localization \cite{sgrignuoli2019localization,razo2024strong}, subdiffusive transport \cite{sgrignuoli2020subdiffusive}, and multifractal field distributions \cite{sgrignuoli2020multifractality}. These unique properties have been successfully exploited in diverse applications ranging from lasing \cite{vardeny2013optics} and optical sensing \cite{razi2019optimization,Lee,Gopinath}, to photodetection \cite{trevino2011circularly} and optical imaging \cite{huang2007optical}, positioning aperiodic systems as a highly promising material platform for the next generation of functional photonic devices.

Within the landscape of deterministic aperiodic optical systems, Vogel spirals stand out for their versatility and their capacity to tune structural order from short-range correlated amorphous configurations to fully uncorrelated random arrangements \cite{trevino2011circularly,dal2012analytical,lawrence2012control,liew2011localized,christofi2016probing,Pollard}. 
In reciprocal space, Vogel spirals are characterized by diffuse circular rings and, in contrast to photonic crystals and quasicrystals, do not have well-defined Bragg peaks \cite{DalNegroCrystals}. The positions of the particles in Vogel spirals are generated in polar coordinates according to $r_{n}=a_{0}\sqrt{n}$, $\theta_{n}=n \alpha$,
where $n$ is a positive integer, $a_{0}$ is the scaling factor determining particle separation, and $\alpha$ is the divergence angle \cite{Naylor,sgrignuoli2019localization,MaciaBook,dal2012analytical,lawrence2012control} that specifies the constant aperture between successive particles in the array. Since $\alpha$ is an irrational angle, Vogel spiral patterns inherently lack both translational and rotational symmetry, setting them apart from periodic crystals and even quasicrystals \cite{DalNegroCrystals}. This absence of symmetry leads to a rich spectrum of structural correlations that can be finely tuned through a single parameter, the divergence angle. As a result, Vogel spirals support an exceptional diversity of electromagnetic resonances, including spatially extended modes, exponentially localized states, and critically confined modes characterized by power-law decay and multifractal intensity distributions \cite{noh2011lasing,mahler2010quasi,prado2021structural}. These complex modal characteristics give rise to unconventional wave transport  phenomena such as localization transitions, subdiffusion, and ultraslow light propagation \cite{sgrignuoli2019localization,razo2024strong,razo2024aperiodicity}, all emerging from deterministic yet aperiodic structural order.
\begin{figure}[!b]
\centering
\includegraphics[width=0.6\textwidth]{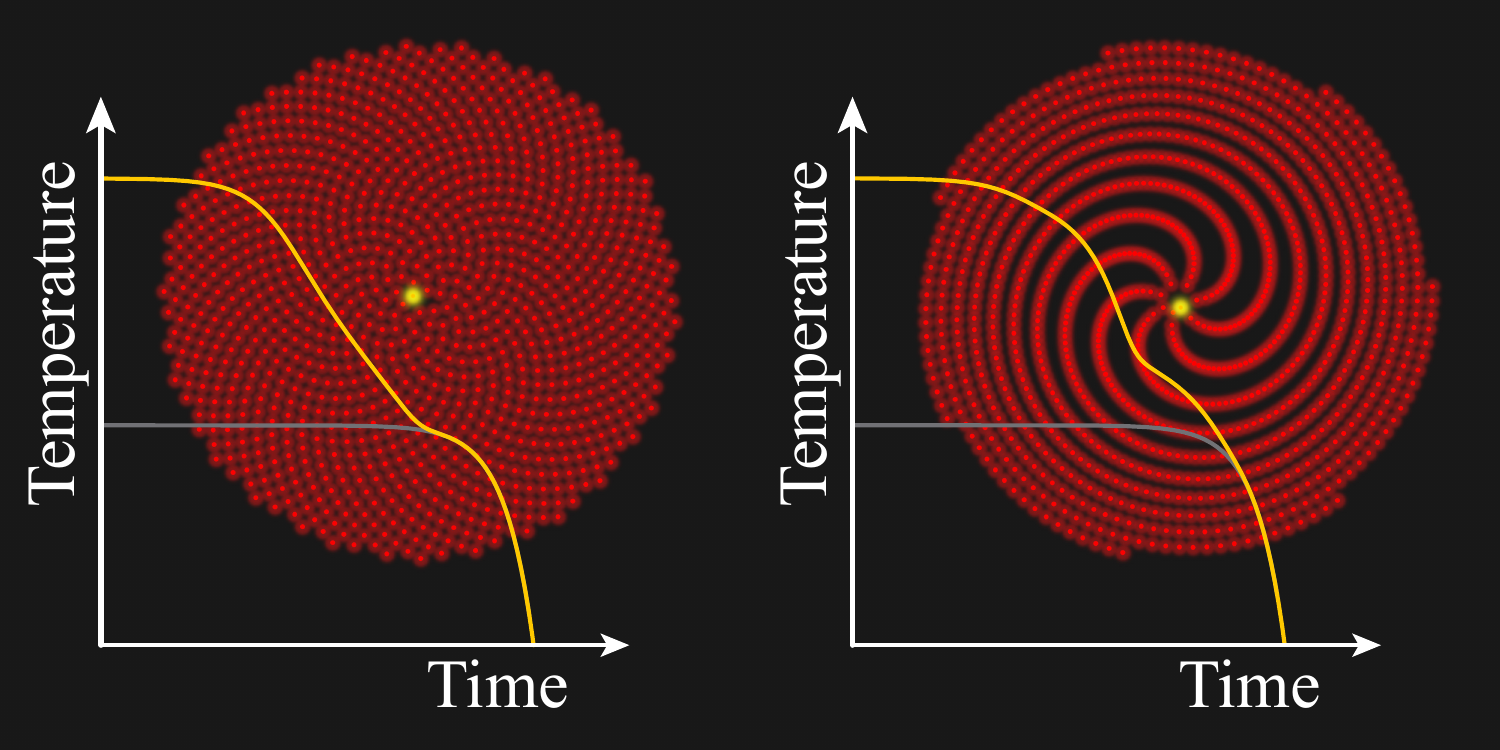}
\caption{Schematic representation of thermalization dynamics in two Voguel spirals where the center particle is initially at a higher temperature than all others in the array.}
\label{fig:schematics}
\end{figure}

Despite extensive exploration of structural correlations in photonic systems that support free-space light propagation, their role in systems governed by surface-confined modes, such as plasmonic and polaritonic materials, remains virtually unexplored. This is a notable gap, as surface polaritons offer a powerful platform for manipulating light at the nanoscale, enabling extreme electromagnetic confinement, large near-field enhancements, and strong light-matter coupling beyond the diffraction limit. These capabilities have driven breakthroughs from subwavelength optics \cite{Novotny_Hecht_2012} to biosensing \cite{biosensor} and quantum nanophotonics \cite{nanocube} to energy harvesting and thermal photonics \cite{Catchpole:08}. However, the influence of deterministic aperiodic order in such systems is still poorly understood. Unveiling how structural correlations shape the modal landscape and energy transport dynamics in polaritonic and plasmonic media is a question of both fundamental and technological importance.
A particularly compelling context to explore these questions is near-field radiative heat transfer (NFRHT), a process intrinsically governed by surface-bound electromagnetic modes \cite{Maier2007}. In the near field, where object separations fall below the thermal wavelength, $\lambda_T = 2\pi\hbar c / (k_B T)$, thermal radiation is no longer described by Planck’s law \cite{narayanaswamy2009breakdown}. Instead, energy exchange is dominated by evanescent surface waves, including surface plasmon and phonon polaritons, leading to heat fluxes that can exceed the blackbody limit by several orders of magnitude \cite{MessinaReview,kim2015radiative,cuevas2018radiative}. While recent studies have investigated the mechanisms of near field interactions in periodic and disordered media \cite{sandersFieldRadiativeHeat2021, Sanders2019, zundel2020active, Latella2018, PhysRevLett.107.114301, PhysRevB.77.075417}, the role of deterministic aperiodic order in controlling how thermal energy flows and thermalizes at the nanoscale remains unexplored. 

In this work, we demonstrate that deterministic aperiodic order serves as a powerful design principle for controlling the dynamics of near-field heat transfer. Using polaritonic nanoparticle ensembles arranged in Vogel spiral geometries, we explore how structural order beyond periodicity or randomness impacts thermal energy flow at the nanoscale. We show that varying the divergence angle---a single geometric parameter that encodes the degree of aperiodic correlation---enables direct control over the temporal dynamics of the thermalization. These results establish deterministic aperiodic structures as a versatile and previously untapped platform for dynamic thermal nanophotonics, opening new avenues for engineering energy transport through correlated light-matter interactions in the near field.

\section{\label{sec:methodology} Methodology}

We analyze near-field radiative heat-transfer (NFRHT) dynamics in deterministic aperiodic arrays of SiC nanoparticles arranged in Vogel spiral geometries. We assume identical spherical nanoparticles with radius $R$ and temperatures $T_i$, embedded in a thermal bath at temperature $T_0 = 300$\,K. As in previous theoretical studies of NFRHT, the nanoparticles are considered suspended in vacuum and the material response of SiC is treated as temperature independent~\cite{MessinaReview, PhysRevLett.107.114301, 10.1063/1.4894622}, so that the radiative heat exchange can be studied independently of conductive or convective channels.
To characterize the thermalization dynamics in these systems, we employ the eigenmode formalism introduced in Ref.~\cite{sandersFieldRadiativeHeat2021}, which naturally incorporates multiple-scattering and many-body effects, and fully accounts for both near- and far-field interactions.

We assume that the nanoparticle radii satisfy $R\ll\lambda_{T_0}$ and that all interparticle separations fulfill $d_{ij}\ge4R$. This lower bound guarantees the validity of the dipole approximation, but the actual separations vary significantly throughout the Vogel spiral, reflecting its aperiodic geometry. In this regime, each nanoparticle located at position $\mathbf{r}_i$ is modeled as a fluctuating electric dipole with polarizability tensor $\boldsymbol{\alpha}_i(\omega)$ dominated by optical phonon–polariton resonances, while magnetic dipole and higher-order multipole contributions can be safely neglected. Indeed, whereas the magnetic polarizability scales as $R^5/\lambda^2$, the electric one scales as $R^3$. Consequently, the ratio of magnetic to electric dipole strengths decreases as $(R/\lambda)^2$, rendering the magnetic term negligible in the near- and mid-infrared regimes relevant for the nonmagnetic polar dielectric nanoparticles considered here. Although not treated in this work, the formalism can be generalized to metallic nanoparticles by including a magnetic polarizability, since in that case magnetic dipole moments generated by eddy currents can become comparable to or even exceed the electric contribution~\cite{PhysRevB.86.075466,PhysRevB.95.125411}. Such effects, however, are negligible for SiC and similar polar dielectric nanoparticles with the sizes considered here.

Within fluctuational electrodynamics, the instantaneous power absorbed by particle $i$ is 
\begin{equation}
P_i(t)=\Big\langle \mathbf{E}_i(t)\cdot\frac{\partial \mathbf{p}_i(t)}{\partial t}\Big\rangle = -\int_{-\infty}^{\infty}\!\frac{d\omega\,d\omega'}{(2\pi)^2}e^{-i(\omega-\omega')t}\, i\omega\,\big\langle \mathbf{E}_i^*(\omega')\cdot \mathbf{p}_i(\omega)\big\rangle,
\label{power}
\end{equation}
where $\mathbf{E}_i$ and $\mathbf{p}_i$ are, respectively, the total local electric field at $\mathbf{r}_i$ and the dipole moment of particle $i$. Here, the angular brackets denote ensemble averages over thermal fluctuations.  The many-body electromagnetic response is obtained by solving the self-consistent linear system driven by the fluctuating sources in the particles and in the environment. For a collection of $N$ dipoles, the solution to this scattering problem can be written as \cite{MessinaReview, 10.1063/1.4894622,  PhysRevB.88.104307}
\begin{align}
\mathbf{p}_i(\omega) &= 
\sum_{j=1}^{N}\!\!\left[\mathbf{A}_{ij}\,\mathbf{p}_j^{\mathrm{fl}}(\omega)
+ \mathbf{B}_{ij}\,\mathbf{E}_j^{\mathrm{fl}}(\omega)\right], \label{pi_full}
\end{align}
\begin{align}
\mathbf{E}_i(\omega) &= 
\sum_{j=1}^{N}\!\!\left[\mathbf{C}_{ij}\,\mathbf{p}_j^{\mathrm{fl}}(\omega)
+ \mathbf{D}_{ij}\,\mathbf{E}_j^{\mathrm{fl}}(\omega)\right], \label{Ei_full}
\end{align}
with $3N\times 3N$ block matrices 
$\mathbf{A}=[\mathbf{I}-\boldsymbol{\alpha}\mathbf{G}]^{-1}$
$\mathbf{B}=\mathbf{A}\boldsymbol{\alpha}$,
$\mathbf{C}=(\mathbf{G}+\mathbf{G}^0)\mathbf{A}$, 
and $\mathbf{D}=\mathbf{I}+\mathbf{C}\boldsymbol{\alpha}$,  which “dress’’ the fluctuating sources, fully accounting for multiple scattering and many-body coupling within the electric dipole approximation.
Here, $\boldsymbol{\alpha}=\mathrm{diag}(\alpha_1\mathbf{I}_{3\times 3},\ldots,\alpha_N\mathbf{I}_{3\times 3})$ is the block-diagonal polarizability, where $\mathbf{I}_{3\times 3}$ is the $3\times3$ identity matrix, and $\mathbf{G}^0=(2i\omega^3/3c^3)\mathbf{I}$. Finally, 
$\mathbf{G}$ is the free-space dipole–dipole interaction tensor with components \cite{Novotny_Hecht_2006} 
\begin{equation}
\mathbf{G}_{ij}=\frac{e^{ikd_{ij}}}{d_{ij}^3}
\left[(k^2d_{ij}^2+ikd_{ij}-1)\mathbf{I}_{3\times 3}-(k^2d_{ij}^2+3ikd_{ij}-3)\,\hat{\mathbf{d}}_{ij}\hat{\mathbf{d}}_{ij}\right],\nonumber
\end{equation}
for $i\neq j$, and zero for $i = j$. 
$k=\omega/c$, $\hat{\mathbf{d}}_{ij}=(\mathbf{r}_i-\mathbf{r}_j)/d_{ij}$.

As in previous studies of near-field radiative dynamics, we assume local thermal equilibrium for each nanoparticle. Substituting Eqs.~\ref{pi_full} and~\ref{Ei_full} into Eq.~\ref{power}, making use of the fluctuation–dissipation theorem~\cite{Rytov1959}
\begin{align}
\big\langle \mathbf{p}_i^{\mathrm{fl}}(\omega)\mathbf{p}_j^{\mathrm{fl}\dagger}(\omega')\big\rangle 
&=4\pi\hbar\,\delta_{ij}\delta(\omega-\omega')\,\mathrm{Im}\{\boldsymbol{\chi}_i\}\,\Big[n(\omega,T_i)+\frac{1}{2}\Big], \nonumber \\
\big\langle \mathbf{E}_i^{\mathrm{fl}}(\omega)\mathbf{E}_j^{\mathrm{fl}\dagger}(\omega')\big\rangle 
&=4\pi\hbar\,\delta(\omega-\omega')\,\mathrm{Im}\{\mathbf{G}_{ij}+\mathbf{G}^0_{ij}\}\,\Big[n(\omega,T_0)+\frac{1}{2}\Big], \nonumber
\end{align}
where  $n(\omega,T)=[\exp(\hbar\omega/k_BT)-1]^{-1}$ is the Bose–Einstein distribution and $\boldsymbol{\chi}=\boldsymbol{\alpha}-\mathbf{G}^0\boldsymbol{\alpha}^\dagger\boldsymbol{\alpha}$ is the dressed susceptibility, and enforcing that $P_i=0$ when all $T_j=T_0$, we derive
\begin{equation}
P_i=\sum_{j=1}^N\int_0^\infty d\omega\, f_{ij}(\omega)\,[n(\omega,T_j)-n(\omega,T_0)], \nonumber
\end{equation}
where the coupling kernel is
\begin{equation}
f_{ij}(\omega)=\frac{2\hbar\omega}{\pi}\,\mathrm{Tr}\!\left\{\mathrm{Im}\big[\mathbf{A}_{ij}\big]\ \mathrm{Im}\big[\boldsymbol{\chi}_j\big]\ \mathbf{C}_{ij}^\dagger\right\},\nonumber
\end{equation}
and the trace is taken over Cartesian components.

\begin{figure}[!t]
\centering
\includegraphics[width=0.6\textwidth]{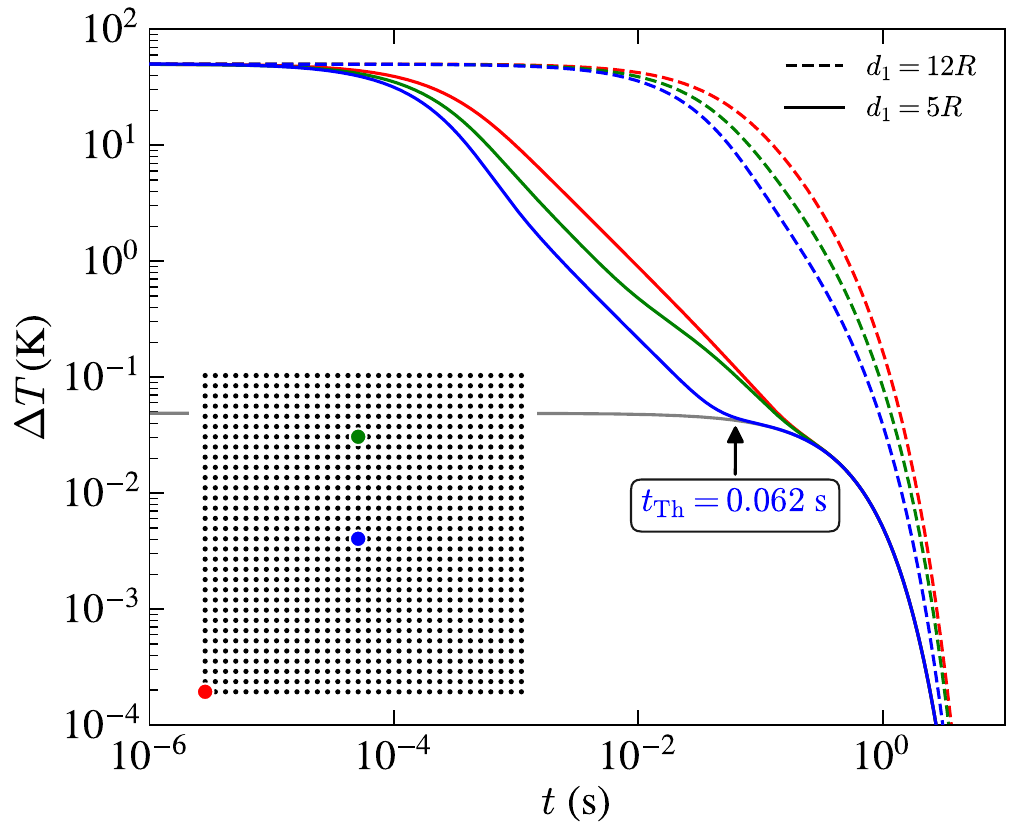}
\caption{Thermalization dynamics for a square array of $N = 1024$ SiC nanoparticles with radius $R = 25\,$nm and interparticle distance $d_1$, under different initial conditions. Colored curves show the time evolution of the temperature of the selected nanoparticle that begins at $\Delta T = 50\,$K while all others are initially in thermal equilibrium with the environment at  $T_0 = 300\,$K. The inset shows the position of the initial hot particles: blue (red) represents the closest (farthest) point from the geometrical center of the array, and the green point is at the median distance. The gray curve corresponds to the case where all nanoparticles are initially at $ \Delta T = 50/1024\,$K and the arrow indicates the interparticle thermalization time $t_\mathrm{Th}$ for the initial condition depicted in blue with $d_1 = 5R$.}
\label{fig:square}
\end{figure}

The temperature dynamics of the nanoparticles is determined by the ratio between the power they absorb $P_i$ and their heat capacities $C_i$ according to $C_i\,\dot{T}_i=-P_i$.  We assume a separation of time scales between the fast electromagnetic response and the slower thermal evolution.
Under this adiabatic approximation, the radiative power at each time step is evaluated using equilibrium response functions at $T_0$, while the resulting power determines how the particle temperatures evolve in time. Writing $T_i=T_0+\Delta T_i$ and linearizing the Bose–Einstein function around $T_0$ leads to the following differential equation governing the time evolution of the system:
\begin{equation}
\frac{d}{dt}\,\Delta\mathbf{T}(t)=-\mathbf{H}\,\Delta\mathbf{T}(t),\qquad \mathbf{H}=\boldsymbol{\Gamma}^{-1}\mathbf{F},
\label{DeltaT}
\end{equation}
where $\Delta\mathbf{T}=(\Delta T_1,\ldots,\Delta T_N)^\top$, $\boldsymbol{\Gamma}=\mathrm{diag}(C_1,\ldots,C_N)$, and the symmetric coupling matrix $\mathbf{F}$ is
\begin{equation}
F_{ij}=-\int_0^\infty d\omega\, f_{ij}(\omega)\Big(\frac{\partial n}{\partial T}\Big)_{T_0}.\nonumber
\end{equation}
Because the frequency integrals are evaluated at the bath temperature $T_0$ and the dielectric response of SiC is treated as temperature independent, the radiative coupling matrix $\mathbf{F}$ (and thus $\mathbf{H}$) is time independent.
All temporal evolution arises solely from the temperature deviations $\Delta T_i(t)$. This linearization approach is controlled and accurate for moderate temperature excursions and mid-IR resonances. A systematic estimate of the truncation error obtained by expanding the difference $n(\omega,T_j)-n(\omega,T_0)$ shows that the relative error scales as
\begin{equation}
\varepsilon \simeq \frac{1}{2}\frac{\Delta T}{T_0}
\left[
\frac{\hbar\omega_0}{k_BT_0}\coth\!\Big(\frac{\hbar\omega_0}{2k_BT_0}\Big) -2
\right],\nonumber
\end{equation}
where $\omega_0$ is the characteristic material resonance that dominates the spectral integrals entering $f_{ij}$. In practice, one can numerically show that for $\max_i|\Delta T_i|/T_0\lesssim 1/3$ and $\hbar\omega_0\ll k_BT_0$ (as in SiC around its phonon–polariton band), the eigenmode predictions agree closely with the fully nonlinear evaluation of $P_i$~\cite{sandersFieldRadiativeHeat2021}. All the calculations presented in this work satisfy these validity conditions.

The collective thermalization behavior of the system is fully determined by the properties of the matrix $\mathbf{H}=\boldsymbol{\Gamma}^{-1}\mathbf{F}$. This matrix governs the coupling between temperature deviations of different particles and encapsulates both their radiative interactions (through $\mathbf{F}$) and their thermal inertias (through $\boldsymbol{\Gamma}$). When all nanoparticles have the same heat capacity, $\mathbf{H}$ is symmetric, real, and positive definite, and its spectral decomposition follows directly from the standard eigenvalue theorem. However, when the heat capacities differ, $\mathbf{H}$ is no longer symmetric because $\boldsymbol{\Gamma}^{-1}$ and $\mathbf{F}$ do not commute. Despite this asymmetry, $\mathbf{H}$ remains diagonalizable and possesses a complete set of real eigenvalues and eigenvectors (see \cite{sandersFieldRadiativeHeat2021} for details).
The solution of the linear system in Eq.~\ref{DeltaT} can therefore be expressed as a superposition of exponentially decaying thermal eigenmodes,
\begin{equation}\label{eq:modal_decomp}
\Delta\mathbf{T}(t)=\sum_{\mu=1}^{N} a_\mu\,e^{-\lambda_\mu t}\,\Delta\mathbf{T}^{(\mu)}, \nonumber
\end{equation}
where the coefficients $a_\mu$ are determined by the initial temperature distribution $\Delta\mathbf{T}(0)$ via
\begin{equation}
a_\mu=\sum_{i=1}^{N} C_i\,\Delta T_i(0)\,\Delta T_i^{(\mu)}. \nonumber
\end{equation}

Each eigenvalue $\lambda_\mu$ defines a characteristic decay rate, and the corresponding eigenvector $\Delta\mathbf{T}^{(\mu)}$ describes the spatial pattern of temperature deviations associated with that mode. The slowest-decaying mode ($\lambda_1$) represents the collective, uniform equilibration of the ensemble with the environment, while higher-order modes ($\mu>1$) capture faster, spatially structured relaxation processes arising from near-field interactions among the nanoparticles \cite{sandersFieldRadiativeHeat2021}. For numerical evaluation, the spectral kernel $f_{ij}(\omega)$ is computed on an adaptive frequency grid that resolves the phonon–polariton resonance, and the integrals defining $F_{ij}$ are performed by adaptive quadrature until convergence. The matrix $\mathbf{H}$ is assembled and diagonalized with standard routines to obtain decay rates $\lambda_\mu$ and eigenvectors $\Delta\mathbf{T}^{(\mu)}$. This workflow scales efficiently to Vogel spiral arrays with hundreds-to–thousands of particles, enabling systematic exploration of how deterministic aperiodic correlations reshape the modal spectrum and hence the transient NFRHT dynamics.
\vspace{20pt}
\section{\label{sec:sec3} Results and Discussion}
%

To establish a reference for comparison with the results we later derive for Vogel spirals, we first analyze the thermalization dynamics of a periodic $32 \times 32$ square array of $N = 1024$ nanoparticles, each with radius $R = 25\,$nm and interparticle spacing $d_1$. We consider the case in which one of the nanoparticles is initially at $\Delta T= 50\,$K and all others are in thermal equilibrium with the environment at $T_0 = 300\,$K. Figure~\ref{fig:square} shows the evolution of the temperature for different selections of the initial hot nanoparticle. 
We note that when heat is deposited into a single particle, its thermalization dynamics is influenced by the surrounding environment and the available fast-decaying modes that can be excited under the chosen initial condition. 
Thermalization modes involving mostly the nearest neighbors of the initial hot particle dominate the dynamics at first~\cite{sandersFieldRadiativeHeat2021}. Then, heat diffuses throughout the array progressively as this process is propagated to the next nearest neighbors.
The apparent differences observed at early times between the three cases considered stem solely from how each initial condition projects onto the same set of thermal eigenmodes. The eigenmodes and their decay rates are fixed by the geometry and do not depend on which particle is initially heated. Eventually, all particles reach the same temperature $T>T_0$ and the collective temperature of the entire array ultimately decays exponentially with a time constant $1/\lambda_1$.
This final stage is associated with the slowest decay eigenmode $\mu = 1$, which is always involved as long as the total initial heat stored in the array is different from zero\cite{sandersFieldRadiativeHeat2021}. 
In other words, the final thermalization dynamics remains unchanged if, instead, we consider the initial condition in which the heat is equally distributed among all the nanoparticles, as illustrated by the gray curve in Fig.~\ref{fig:square}. Therefore, we define a characteristic interparticle thermalization time $t_\mathrm{Th}$ beyond which thermalization within the array is considered complete and the dynamics are governed by ${\Delta T (t > t_\mathrm{Th})\sim e^{-\lambda_1 t}}$. 
\begin{figure}[!b]
\centering
\includegraphics[width=0.8\textwidth]{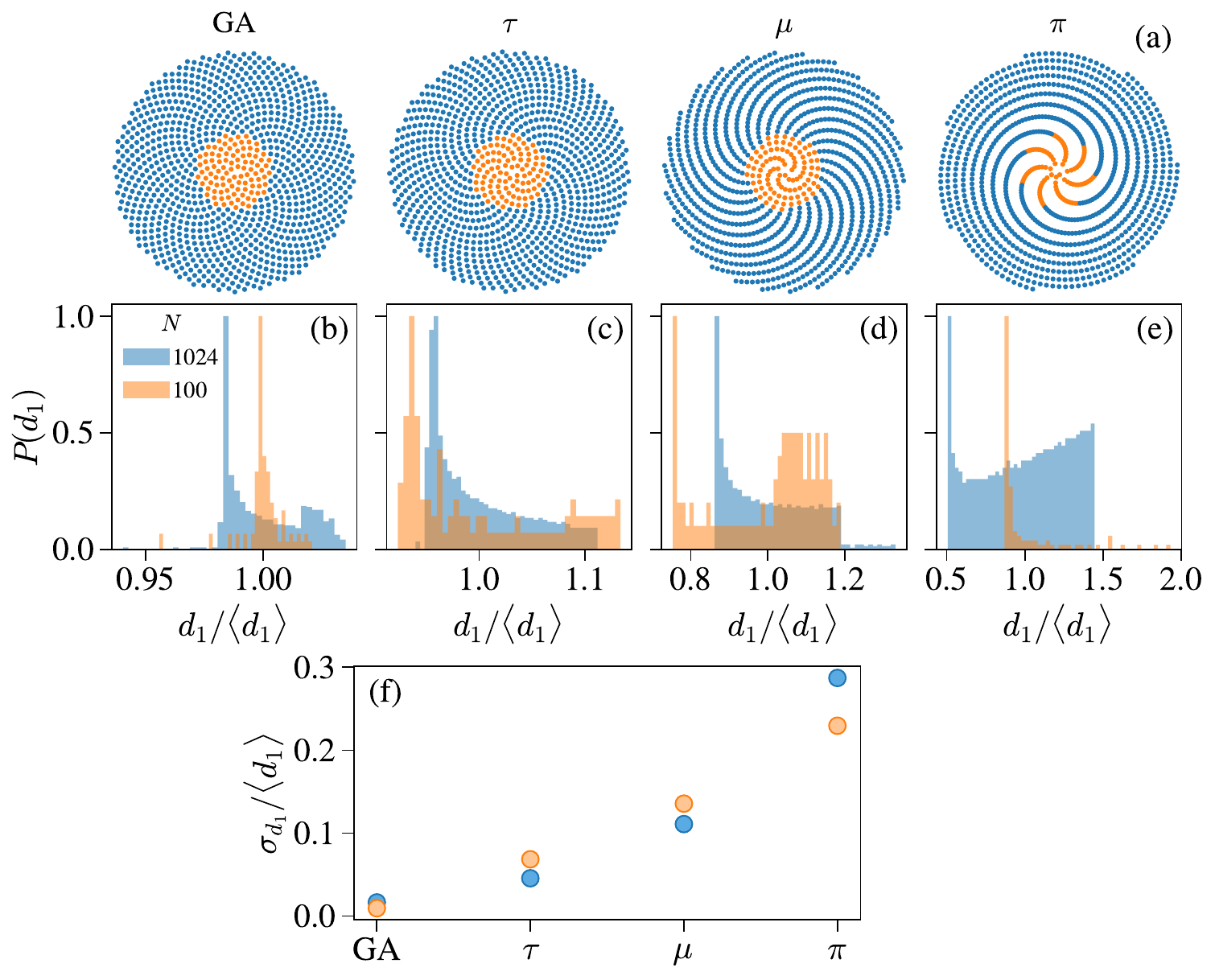}
\caption{(a) Schematic representation of the different Vogel spirals under consideration. (b-e) Distribution of the interparticle distances for the GA (b), $\tau$ (c), $\mu$ (d), and $\pi$ (e) spirals, respectively. Orange and blue dots correspond to the systems with $N=100$ and $N=1024$ nanoparticles, respectively. (f) Standard deviation of the interparticle distances for the Vogel spirals in (b)-(e).}
\label{fig:histograms}
\end{figure}

Indeed, the intricacies of the transient thermalization dynamics are encoded in the values of $t_\mathrm{Th}$ obtained under different initial conditions. 
For $d_1 = 5R$ (solid colored curves in Fig.~\ref{fig:square}),
when the heat is initiated at the center of the square array (blue curve) $t_\mathrm{Th} \approx 0.06$ s, which increases considerably when instead the hot particle is farther from the center (green and red curves). Furthermore, when the interparticle distance is increased to $d_1 = 12R$ (dashed colored curves in Fig.~\ref{fig:square}), the temperature evolution is only slightly different from the one described by $ e^{-\lambda_1 t}$ expected at the final stage of thermalization. Accordingly, $t_\mathrm{Th}$ is 3 orders of magnitude larger in this case. This highlights that radiative dynamics are strongly affected by how particles are arranged in space and therefore can be engineered by introducing structural correlations.
\begin{figure}[!t]
\centering
\includegraphics[width=0.8\textwidth]{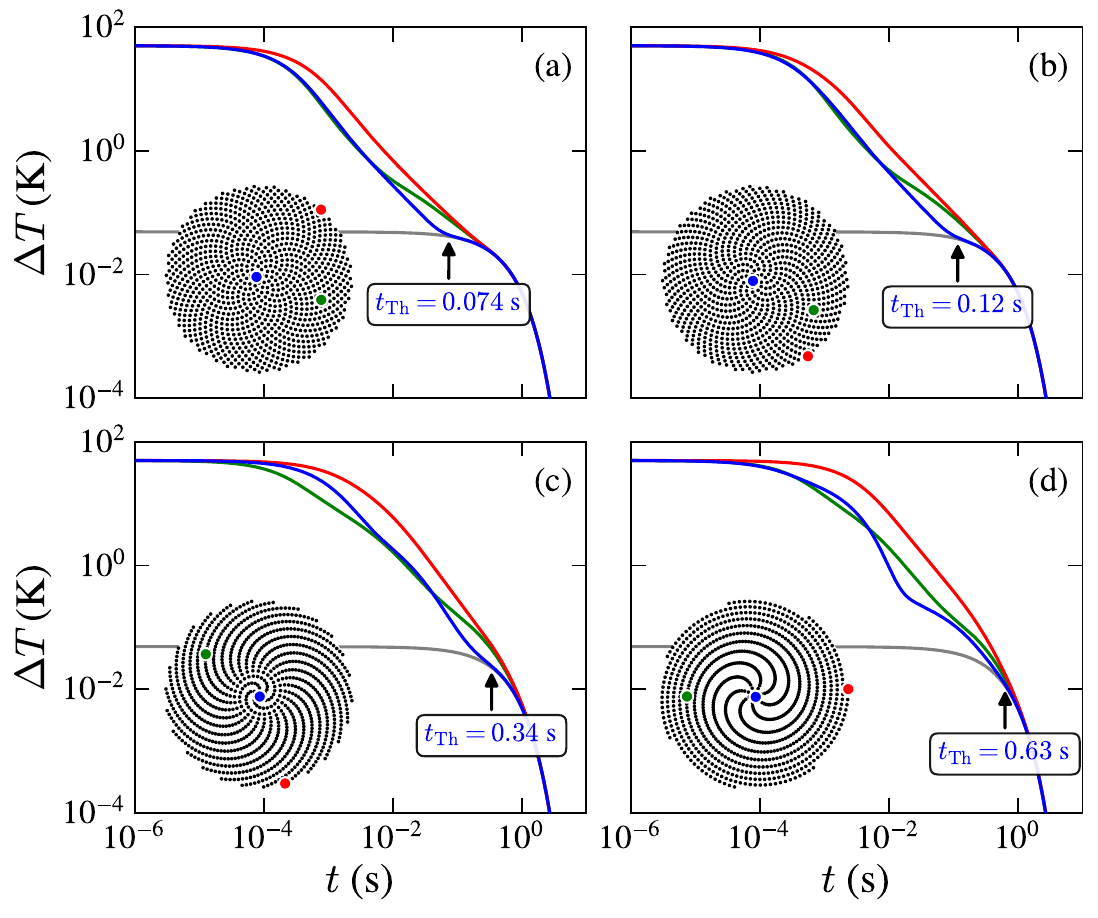}
\caption{Thermalization dynamics for the GA (a), $\tau$ (b), $\mu$ (c), and $\pi$ (d) spirals with $\langle d_1 \rangle = 5R$, under the same conditions as in Fig.~\ref{fig:square}. Insets show the corresponding spiral geometries.}
\label{fig:spirals}
\end{figure}

Next, we analyze the four Vogel spiral geometries under the same conditions. Each spiral is characterized by a divergence angle $\alpha$, derived from an irrational number $\xi$ through the relation $\alpha = 2\pi [1 - \textrm{frac}(\xi)]$, where $\textrm{frac}(\xi)$ denotes the fractional part of $\xi$~\cite{dal2022waves}. In this work, we consider four types of Vogel spirals characterized by $\xi = (1 + \sqrt{5})/2$,  $\xi = (2 + \sqrt{8})/2$, $\xi = (5 + \sqrt{29})/2$, and $\xi = \pi$. These correspond to the golden angle (GA), $\tau$, $\mu$, and $\pi$ spirals, respectively, whose spatial distributions are shown in Fig.~\ref{fig:histograms}(a).  Each geometry yields a distinct distribution of interparticle distances. In Figs.~\ref{fig:histograms}(b)-(e), we present normalized histograms $P(d_1)$ of the interparticle distances $d_1$ for each of the different spirals. The GA and $\tau$ spirals exhibit narrow, peaked distributions, reflecting strong short-range correlations, while the $\mu$ and $\pi$ spirals display broader, more irregular distributions. 
The non-uniform character of a Vogel spiral array is visible from changes in the structural properties of its core (depicted in yellow in Fig.~\ref{fig:histograms}) in comparison to the ones of the larger structure as a whole (shown in blue). This is expected to have an effect on the thermalization dynamics, since they are governed by the environment of where the heat is initiated.
Moreover, the four types of spirals studied here span a significant range of structural order, with the GA resembling periodic arrangements and the $\pi$ being closer to fully uncorrelated systems.
In the following, we set $\langle d_1 \rangle = 5R$ and hence the degree of disorder is quantified via the standard deviation $\sigma_{d_1}$ of $d_1$, which increases monotonically from GA to $\pi$ spirals, as seen in Fig.~\ref{fig:histograms}(f).
 
Figure~\ref{fig:spirals} shows the temperature decay of selected nanoparticles in GA, $\tau$, $\mu$, and $\pi$ spirals. As in the square periodic array analyzed in Fig.~\ref{fig:square}, the final stage of thermalization is governed by $\lambda_1$, which is determined by the geometrical and material properties of the nanoparticles and is largely unaffected by the array geometry \cite{sandersFieldRadiativeHeat2021}. Nonetheless, we observe that the intricate transient dynamics is substantially influenced by the introduction of aperiodic order. 
For instance, it is noticeable from the initial condition depicted in blue how the evolution up to $t_\mathrm{Th}$ in the $\pi$ spiral has much more features than in the case of the GA geometry. This reflects the more irregular local environment of the former array, whose central particle has less neighbors within its reach to transfer heat efficiently through the near-field.
Accordingly, the interparticle thermalization time captures the dependence on the degree of structural order: the GA spiral shows the fastest route to equilibrium ($t_\mathrm{Th} = 0.074\,$s when the heat is initially deposited at the center particle), while the $\pi$ spiral exhibits the slowest ($t_\mathrm{Th} = 0.63\,$s). These differences highlight the role of aperiodic order in shaping collective modal behavior. 
\begin{figure}[!t]
\centering
\includegraphics[width=0.8\textwidth]{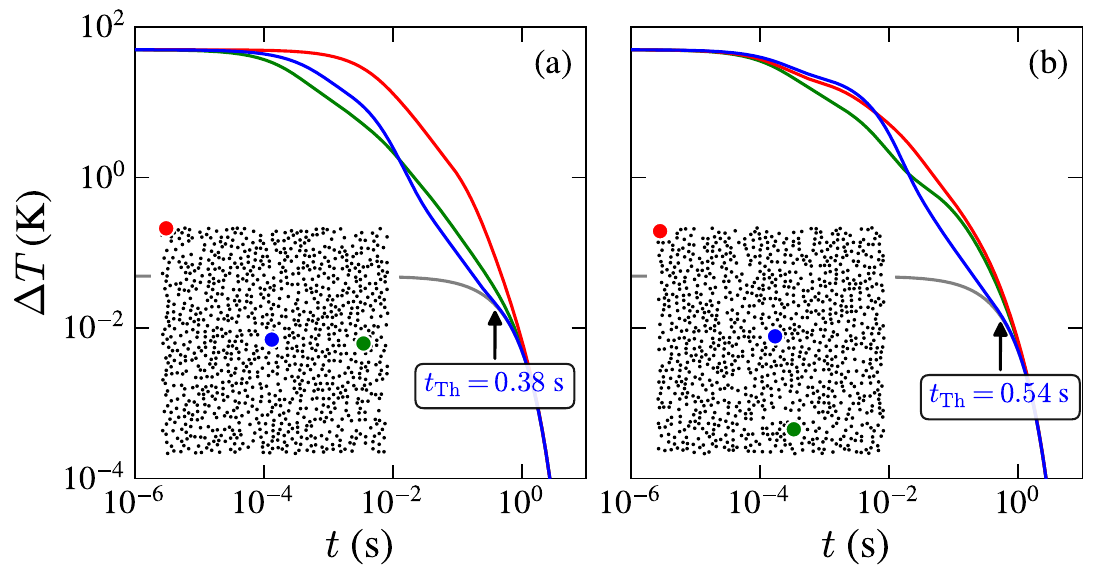}
\caption{Thermalization dynamics for two random arrays of $N=1024$ nanoparticles with constrained $d_\mathrm{min} = 4R$ and $\langle d_1 \rangle = 5R$,  under the same conditions as in Fig.~\ref{fig:square}. Panels (a) and (b) show the fastest and slowest realizations, respectively. Insets show the corresponding geometries.}
\label{fig:random}
\end{figure}

The ability to tailor transient dynamics can be understood through the mechanism revealed by the framework introduced in Ref.~\cite{sandersFieldRadiativeHeat2021}. When radiative heat transfer is dominated by the near field, the modal decomposition in Eq.~\ref{eq:modal_decomp} shows that thermalization in a nanoparticle ensemble proceeds progressively, with an increasing number of nearest neighbors of the initially hot particle becoming involved at each stage. Interparticle distances $d_1$ are therefore a key factor, and Vogel spirals owe their capacity to tune heat transport to the range of structural correlations they can achieve, as illustrated in Fig.~\ref{fig:histograms}.
\begin{figure}[!b]
\centering
\includegraphics[width=0.8\textwidth]{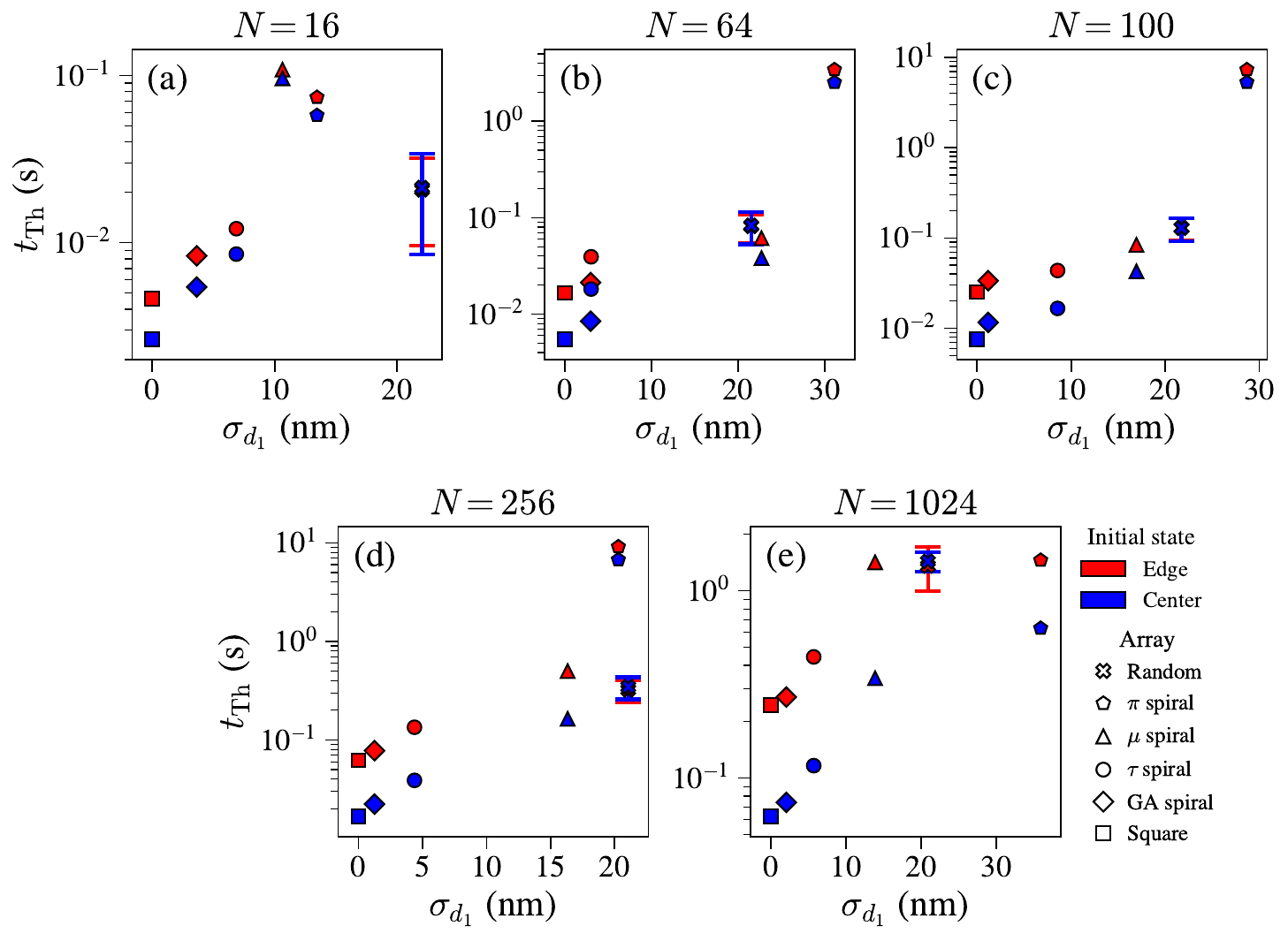}
\caption{Thermalization time $t_\mathrm{Th}$ as a function of the standard deviation of the interparticle distances $\sigma_{d_1}$ for arrays with (a) $N = 16$, (b) $N = 64$, (c) $N = 100$, (d) $N = 256$,  and (e) $N = 1024$ nanoparticles. Blue (red) points correspond to initial conditions with heat at the center (edge) of the array. For random arrays, $t_\mathrm{Th}$ is averaged over $2^{14}/N$ independent configurations and the error bars correspond to standard deviation.}
\label{fig:tth_vs_sd}
\end{figure}

To place our results in context, we also examine the thermalization dynamics in random arrays. They are generated by uniform sampling of points in the plane under the constraints of ${\langle d_1 \rangle = 5R}$ and minimum interparticle distance $d_{\min} \geq 4R$. Because the random ensembles are statistically homogeneous under fixed average spacing and minimum-distance, their structural statistics—, and thus $\sigma_{d_1}$, are essentially independent of $N$. Figure~\ref{fig:random} shows representative dynamics for configurations yielding the shortest and longest $t_\mathrm{Th}$ among the 16 sampled realizations. Although some random arrays perform comparably to the $\mu$ spiral, none outperform GA or $\tau$ spirals, confirming that deterministic aperiodicity can surpass uncorrelated disorder in optimizing NFRHT dynamics.

Finally, we establish a direct link between structural correlations and thermalization dynamics by plotting $t_\mathrm{Th}$ as a function of $\sigma_{d_1}$ in Fig.~\ref{fig:tth_vs_sd} for Vogel arrays of various sizes. The results reveal a robust monotonic relationship: higher structural disorder leads to slower thermalization. Notably, Vogel spirals provide a continuous and tunable interpolation between order and disorder through the divergence angle, offering a unique control parameter for near-field transport engineering.

These findings collectively demonstrate that deterministic aperiodic order—specifically in Vogel spiral configurations—offers a powerful and versatile material platform to modulate the dynamics of radiative heat transfer. In particular, the ability to predictably tune the radiative heat transport characteristics through geometric design alone opens new avenues for developing dynamic thermal photonic materials and devices. While our analysis considers idealized geometries, practical implementations will inevitably present fabrication tolerances, size and position disorder, or surface roughness, which can slightly modify modal lifetimes and spectral features. Nevertheless, the theoretical trends reported here provide clear design guidelines for exploiting deterministic aperiodic order to control near-field radiative heat-transfer dynamics.

\section{\label{sec:sec4} Conclusions}

In this work, we have investigated the role of structural correlations on the near-field radiative heat transfer dynamics in ensembles of polaritonic nanoparticles. Specifically, we have considered array geometries based on aperiodic deterministic Vogel spirals, which enable a controllable interpolation between periodic and random structures. In the transient regime, where near-field radiative heat transfer dominates, we have demonstrated that a higher degree of structural disorder results in slower thermalization. Furthermore, due to the tunable range of aperiodic order offered by Vogel spirals, these structures serve as photonic platforms that enable predictive control over radiative heat transport. These insights pave the way for the design of advanced thermal management systems and nanoscale devices where precise control over energy flow is essential.

\section{Acknowledgments}
M. Prado and F. A. Pinheiro acknowledge financial support from the Brazilian  agencies CAPES, CNPq, and FAPERJ. A. Manjavacas acknowledges support by Grant No.~PID2022-137569NB-C42 funded by MICIU/AEI/10.13039/501100011033 and FEDER, EU. W. J. M. Kort-Kamp acknowledges support by the Laboratory Directed Research and Development program of Los Alamos National Laboratory under Project Number 20240037DR. 
 
\bibliographystyle{iopart-num}
\bibliography{refs.bib}

\end{document}